# Measuring the Dzyaloshinskii-Moriya interaction of the epitaxial Co/Ir(111) interface

Fabian Kloodt-Twesten[1], Susanne Kuhrau[1], Hans Peter Oepen[1], and Robert Frömter[1]

[1] *Universität Hamburg, Center for Hybrid Nanostructures, Luruper Chaussee 149, 22761 Hamburg, Germany*

- - - Abstract - - -

The in-plane orientation of the magnetization in the center of domain walls is measured in Co/Ir(111) as a function of Co thickness via scanning electron microscopy with polarization analysis. Uncapped, thermally evaporated cobalt on an Ir(111) single-crystal surface is imaged *in situ* in ultra-high vacuum. The initial pseudomorphic growth with an atomically flat interface of cobalt on iridium ensures comparability to theoretical calculations and provides a study of an interface that is as ideal as possible. Below a cobalt thickness of 8.8 monolayers, the magnetic domain walls are purely Néel oriented and show a clockwise sense of rotation. For larger thicknesses the plane of rotation changes and the domain walls show a significant Bloch-like contribution, allowing to calculate the strength of the Dzyaloshinskii-Moriya interaction (DMI) from energy minimization. From the angle between the plane of rotation and the domain-wall normal an interfacial DMI parameter $D_\text{s} = -(1.07 \pm 0.05)\,\text{pJ/m}$ is determined, which corresponds to a DMI energy per bond between two Co atoms at the interface of $d_\text{tot} = -(1.04 \pm 0.05)\,\text{meV}$.

## 1. Introduction

The Dzyaloshinskii-Moriya interaction (DMI) [1, 2] is an antisymmetric exchange interaction that occurs due to the lack of inversion symmetry in a crystal structure (bulk type) or at the interface between two different materials (interface type). Caused by the spin-orbit interaction, the interfacial DMI reduces the energy of 180° domain walls between out-of-plane magnetized domains by initiating a laterally rotating spin configuration with a fixed sense of rotation. As a consequence, the domain-wall energy can even become negative in suitable materials, which results in a spin-spiral or skyrmion phase [3-7]. Recently, layered systems with a stacking of iridium, cobalt and platinum have drawn a lot of attention [8-18] as they offer the possibility to stabilize magnetic skyrmions at room temperature (RT) [13]. A *b-initio* calculations of the ideally ordered Co/Ir(111) interface predict [8, 9, 19, 20] that iridium induces a DMI of opposite sense of rotation of magnetization compared to platinum, resulting in a higher absolute DMI strength when cobalt is sandwiched between both materials. Unfortunately, the calculations yield conflicting results for the expected strength of the DMI. Such calculations are very sensitive to details of the band structure [20, 21], the hybridization of the materials at the interface [20-22], the lattice-relaxation procedure, as well as the stacking of the lattice [23] and the assumed energy functional (i.e., considering first order in spin-orbit coupling [8, 20, 24, 25], or accounting only for nearest-neighbor DMI [19, 23]). Furthermore, extracting the DMI from a spin-spiral state makes it necessary to consider a reasonably small wave vector [21, 25, 26] ($q \to 0$ in the micromagnetic limit) and hence calculations might result in material-dependent errors when considering the energy of 90° spin-spirals as in [9, 19]. Despite these various sources of error, little is known about the actual reliability of the individual calculations.

Unfortunately, the authors of the three independent theoretical studies existing for Co/Ir(111) use different units and definitions, and do not compare their results. So, in the first place, a common description is required. Even though the DMI is an interface property, in micromagnetic calculations of thin, rigidly coupled films it is often treated as a scaled bulk property $D_\text{s}/t$ where $t$ is the thickness of the ferromagnetic material and $D_\text{s}$ is the interfacial DMI constant [10, 27-29]. In the following, $D_\text{s}$ will be used for comparison between the different results. Definitions and conversions can be found in the Supplemental Material S1 [30] and were approved by the corresponding authors. Yang *et al.* investigated a bilayer system of 3 monolayers (ML) cobalt on 3 ML Ir(111) [9, 19] and obtain $D_\text{s} = -0.42$ pJ/m. Perini *et al.* calculated the DMI of 1 ML cobalt on 9 ML of Ir(111) and report $D_\text{s} = -1.10$ pJ/m [8]. Belabbes *et al.* give the highest DMI of $D_\text{s} = -7.66$ pJ/m calculating 1 ML of cobalt on 6 ML of Ir(111) [20]. Even though all three published calculations agree in sign, they differ by more than an order of magnitude in size. This large discrepancy cannot be explained by the different amount of simulated cobalt layers, as Yang *et al.* [9, 19] showed that the DMI depends particularly on the first monolayer and the contributions of the next layers are of minor relevance, resulting in the $1/t$ dependence mentioned above.

Precise measurements of the DMI in a system comparable to theory are necessary to give insight into its microscopic origin and lead towards more reliable calculations. As the DMI is very sensitive to the interface quality, epitaxial films that wet the single-crystal surface and grow in a layer-by-layer fashion are best suited for this kind of investigation and ensure the previously mentioned comparability to theoretical calculations. In the case of Co/Ir(111) the first monolayer of thermally evaporated cobalt has been observed to grow pseudomorphically on the Ir(111) surface without intermixing [8, 31, 32], and an fcc stacking is supposed [8]. Therefore, this close-to-ideal system can be well compared to the system calculated in theory; in contrast to the more often studied sputtered film systems.

The spin-polarized scanning tunneling microscopy study of Perini *et al.* [8] is performed on such a well-ordered interface, as a pseudomorphic cobalt monolayer on an Ir(111) single crystal is investigated. This experiment confirms the negative sign of $D_s$ but gives no value for its strength [8]. Chen *et al.* publish a positive value of $D_s = (0.37 \pm 0.09)$ pJ/m [33] for this interface; their quantitative analysis, however, appears to be based on inaccurate assumptions [34]. Further experimental reports of the DMI strength are only available for sputtered Co/Ir(111) interfaces but, due to the aforementioned problems, they show an even larger spread of values. They propose both negative [12, 13] as well as positive [29, 35, 36] signs of the DMI, which is attributed to different interface properties [35, 36] and is consequently no intrinsic property of the Co/Ir(111) interface. In conclusion, the sign of the DMI of the ideal system may be considered as known. However, the actual strength of the DMI is still an open question.

In the present study, the Dzyaloshinskii-Moriya interaction of the epitaxially grown Co/Ir(111) is measured by imaging the in-plane orientation of the magnetization within 180° domain walls (DWs) via scanning electron microscopy with polarization analysis (SEMPA). As DWs in a single uncovered Cobalt layer are investigated, the measured DMI can be unambiguously assigned to the Co/Ir(111) interface. A DMI contribution from the vacuum/Co interface is not to be expected [37]. In the absence of DMI, the magnetization of a DW rotates in the wall plane, reducing stray-field energy by avoiding magnetic volume charges, which is known as a Bloch wall. An additional interfacial DMI, favoring Néel-oriented DWs, tilts the rotational plane of the moments in the domain wall about the surface normal towards the normal of the DW. Therefore, the angle $\Phi$ of the rotational plane with respect to the DW normal allows for analyzing the energy balance between the DMI and the stray-field energy, as long as both contributions are comparable in size [23, 33, 38- 41]. In an illustrative picture, $\Phi$ is the angle between the in-plane magnetization at the center of a DW with respect to the normal of the DW, which is why we denote it as angle of DW magnetization. $\Phi$ is 0°, 180° for a Néel orientation with clockwise, counterclockwise sense of rotation and −90°, +90° for a Bloch orientation with clockwise, counterclockwise chirality, respectively.

## 2. Experiment

SEMPA is well established as a magnetic imaging technique for the investigation of static magnetic structures ranging from hundreds of microns down to 3 nm in size [42]. It is a surface sensitive technique that measures two components of the magnetization simultaneously [43-45] with a depth of information of about 0.5 nm [42, 46] Recently, also the nanosecond magnetization dynamics became accessible [47-49]. The SEMPA system used here is sensitive to the in-plane spin polarization and allows for vectorial imaging of the in-plane magnetization orientation with 4° precision for each pixel of an image [50], which is prerequisite for the investigation of the angle of DW magnetization [17, 38, 51].

The epitaxial Co/Ir(111) sample was prepared in a preparation chamber (base pressure of $5 \times 10^{-11}$ mbar) directly attached to the SEMPA instrument. A clean, atomically flat Ir(111) surface was obtained by argon ion sputtering and periodic annealing of the bulk single-crystal substrate. The cleanliness of the surface was confirmed by Auger electron spectroscopy (AES). Low-energy electron diffraction (LEED) revealed a sharp p(1 × 1) pattern. Subsequently, a wedge-shaped cobalt film was evaporated at RT. As the uncertainty in cobalt thickness is the leading term in the accuracy of DMI determination, a thorough calibration of the evaporation rate was performed. In this way, an effective "magnetic" thickness is obtained, which makes the analysis less dependent on growth-related variations of film density. The sample preparation is described in detail in Supplemental Material S2 [30].

## 3. Results and discussion

Magnetic domain patterns of several cobalt films were recorded within a thickness range of 3.5 to 17 ML. At lower thicknesses the as-grown cobalt films show an out-of-plane magnetized multi-domain state. At $(9.7 \pm 0.3)$ ML the spin-reorientation transition (SRT) towards easy plane sets in, which proceeds via magnetization canting [52]. Here, higher-order terms of the anisotropy needed to be considered in the analysis [53, 54], which is beyond the scope of this article. Therefore, only data significantly below the onset of the SRT will be analyzed. Starting from 7.1 ML cobalt thickness, DWs could be resolved with sufficient accuracy to extract angles of DW magnetization from the SEMPA micrographs. For cobalt thicknesses below 7.1 ML the DWs were too narrow to be adequately imaged. In the lower thickness range, magnetic domains with out-of-plane orientation could still be imaged by slightly tilting the sample, demonstrating that the cobalt layer is ferromagnetic at least down to 3.5 ML.

The magnetic domain structure of 8.6 ML cobalt is displayed in Fig. 1. Panel (a) and (b) show the vertical and horizontal in-plane component of the magnetization, respectively. A slight sample tilt of 10° about the horizontal axis ensures that a small projection of the out-of-plane component is visible in the vertical component image (a), which allows to derive the sense of rotation in the DWs. Fig. 2(c) gives a full three-dimensional map of the magnetization. The up (black)/down (white) domain contrast is obtained from the contribution of the perpendicular domains observed in (a). The in-plane orientation of the domain-wall magnetization is color coded according to the color wheel in the lower left corner. It is calculated from the individual component images (a) and (b). The data show Néel-oriented DWs with a fixed, clockwise sense of rotation (see sketch for definition). This finding is in agreement with recent SP-STM investigations of the cobalt monolayer on Ir(111) [8] and is evidence of a negative DMI. Further inspection via line profiles in Fig. 1 yields that at this thickness the DWs are purely Néel oriented ($\Phi = 0°$) and thus only a lower threshold for the DMI strength can be derived.

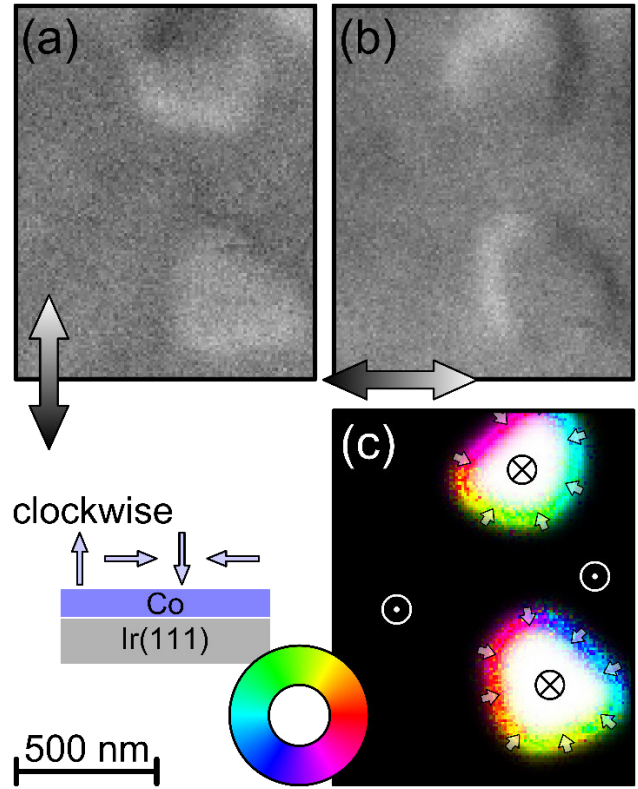

*FIG. 1. Magnetic domain structure of a 8.6 ML cobalt film grown on the Ir(111) single-crystal surface. (a) and (b) display the vertical and horizontal in-plane components of the spin polarization. Due to a sample tilt of 10° about the horizontal axis an additional projection of the out-of-plane magnetization component is visible in the vertical component image (a). From the raw data a complete three-dimensional map of the magnetization vector is assembled in (c). The black and white contrast indicates domains pointing up and down, respectively. The magnetization components in the film plane are color coded according to the given color wheel. Domains pointing into and out of the surface are separated by Néel-oriented domain walls with clockwise sense of rotation.*

A value of the DMI strength can be obtained when the domain wall starts to tilt towards a Bloch wall at higher cobalt thickness. FIG. 2(a) shows a line scan across two DWs at a Co thickness of 9.0 ML. Blue circles and red squares correspond to the horizontal (left axis) and vertical (right axis) component of the in-plane magnetization, respectively. The error bars are calculated based on the Poisson statistics of the individual electron count per channel, as described in [50]. The corresponding SEMPA micrograph in Fig. 2(b) (horizontal component shown) was recorded beforehand and used to select the position of the line scan, which is marked in blue. In Fig. 2(c) a sketch illustrates the definition of the in-plane angle of magnetization within the DW with respect to its normal.

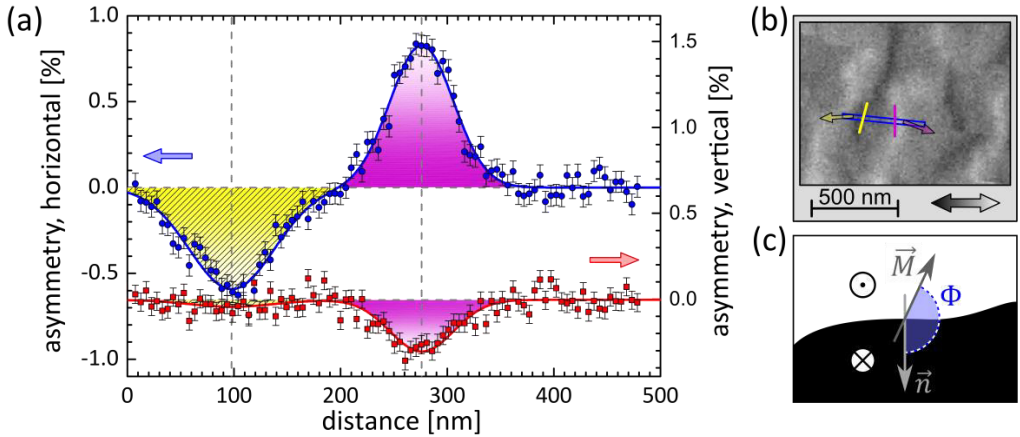

FIG. 2. (a) Line scan across two domain walls at 9.0 ML cobalt thickness. Blue circles and red squares correspond to the horizontal (left axis) and vertical (right axis) components of the in-plane magnetization. In (b) an overview image of the horizontal magnetization component is shown, that was recorded beforehand and used to position the line scan. The DWs marked in yellow and purple correspond to the shading of the Gaussian profiles and the arrows indicate the extracted DW magnetizations. Both DWs are not purely Néel-oriented and show opposite Bloch rotations, yielding angles of DW magnetization of about +/-20°. Panel (c) illustrates the definition of this angle $\Phi$ between the magnetization vector $\vec{M}$ in the center of a DW and the normal $\vec{n}$ of the DW.

As the apparent width of the DWs is obviously dominated by the instrumental resolution, a Gaussian profile is expected that originates from the shape of the primary beam. The profiles of both magnetization components are analyzed in a single fit procedure comprising two independent Gaussian profiles for the two DWs for each component. While the integrated intensities of the horizontal and vertical components are independent parameters for each Gaussian, the standard deviation (or the width) and the position are shared (see Supplemental Material S3 [30]). A remaining contribution from the out-of-plane domains has been included in the fit using a Gaussian error function. This out-of-plane contribution is already corrected in the given data. It corresponds to a 5° tilt for this image, which has only a negligible influence on the in-plane domain contrast [38]. The DW to the right-hand side (marked in purple in the inset) is oriented within an uncertainty of 2° along the image vertical. For this reason, the horizontal magnetization component directly reflects the Néel contribution to the rotation of the DW, while the vertical component gives the Bloch contribution. It is clearly visible that at this investigated thickness the DW is mainly but not purely Néel oriented. From the ratio of the respective integrated intensities of the Gaussians a DW magnetization angle $\Phi = -(20.1 \pm 2.6)°$ is calculated. The situation is different for the DW to the left (yellow) that is rotated by $(15 \pm 5)°$ with respect to the image vertical. Within experimental accuracy the vertical magnetization component is close to zero, which means that the DW magnetization is aligned with the horizontal axis. An analysis of the fit yields an angle of $(18.8 \pm 5.5)°$. The errors for both angles of DW magnetization mainly originate from the uncertainty in determining the local wall orientation and not from the magnetization orientation. Although both angles have opposite signs, they give a consistent picture, as the energy minimization that involves stray-field energy and DMI only defines the absolute value and not the sign of $\Phi$. Contrary to the DMI, which favors only one Néel orientation, both Bloch orientations are degenerate in energy, which is consistent with recent studies on other materials [39-41].

The different widths visible in the Gaussian fits can be attributed to astigmatism of the scanning electron beam, yielding different convolutions depending on the orientation of line-like objects (see Supplemental Material S3 [30]). However, the integral intensities $V$ are not influenced by this effect as they depend only on the actual DW width and the in-plane contrast and make a precise measurement of the angle of DW magnetization possible [38]. In good agreement, one obtains $V = (0.65 \pm 0.03)$ nm for the left DW and $V = (0.61 \pm 0.06)$ nm for the right DW, which leads to a DW width according to the definition by Lilley [55] of $d_\text{w} = (14.6 \pm 0.7)$ nm.

### 4. Determining the DMI strength

By minimizing the energy of a DW including DMI and stray-field energy via micromagnetic modelling, the DMI parameter $D_\text{s}$ can be determined from comparison to the measured equilibrium angle of DW magnetization. The main problem in modelling is to determine the stray-field energy of the magnetic volume charges accurately. This energy can be obtained numerically [33, 39-41] or using an analytical approximation [56-58]. In this study, the measured angles of DW magnetization are matched with micromagnetic simulations of DWs using MicroMagnum [59]. For comparison, also the analytical models are fitted to the experimental data.

In Fig. 3 a plot of the angle of DW magnetization as function of the cobalt thickness is shown (gray circles), together with further data points, that have been obtained in the same way as in Fig 2 at different thicknesses. For lower cobalt thickness the walls show within the error margin an angle of $\Phi = 0°$, corresponding to purely Néel-oriented walls with a clockwise sense of rotation. Above a cobalt thickness of 8.8 ML, DWs that are not purely Néel-oriented are observed. The red squares in Fig. 3 correspond to micromagnetic simulations of DWs using a DMI parameter of $D_\text{s} = -1.065 \frac{\text{pJ}}{\text{m}}$ (details on the simulations in Supplemental Material S5 [30]) In a next step, we varied the value of $D_s$ in the simulations (not shown) to find the uncertainty in fitting our experimental results. We obtain $D_s = -(1.07 \pm 0.01)\frac{\text{pJ}}{\text{m}}$. An additional and independent analysis has been performed by fitting the approximate solution developed by Lemesh *et al.* [58] (details for the analysis in Supplemental Material S4 [30]) with $D_s$ as only fit parameter. This fit yields $D_s = -(1.04 \pm 0.01)\frac{\text{pJ}}{\text{m}}$ and is shown in blue with a blue shade visualizing the confidence interval at 1 sigma. Lemesh *et al.* interpolate between the solutions for ultra-thin (previously given by Tarasenko *et al.* [56]) and thick films giving a better match than the ultra-thin film solution. Using directly the latter, as in [38, 60], results in an overestimation of the DMI strength by 10%.

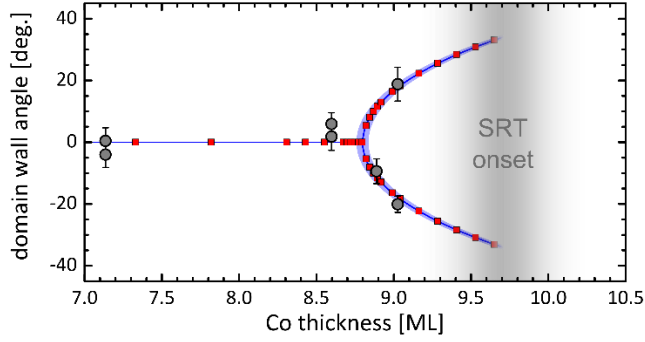

*FIG. 3. Measured angles of the DW magnetization as function of cobalt thickness (gray circles). Below a cobalt thickness of 8.8 ML purely Néel-oriented domain walls are observed that correspond to a clockwise sense of rotation. Above, the domain walls show a significant Bloch contribution until the SRT sets in at $(9.7 \pm 0.3)$ ML. The measured angles match with micromagnetic simulations of DWs using a DMI strength of $D_s = -1.07$ pJ/m (red squares). A fit of the approximating equation of Lemesh et al. [58] results in a similar value of $D_s = -(1.04 \pm 0.01)$ pJ/m (blue line). The blue shaded region visualizes the confidence band at 1 sigma.*

In addition to the small statistical error margin in the DMI parameter $D_s$, a systematic error caused by the thickness calibration of the evaporated Co has to be considered (see Supplemental Material S2). Accounting for 2.1% uncertainty in the thickness calibration, which results from calibration by atomic force microscopy, one obtains:

$$D_\text{s} = -(1.07 \pm 0.05)\frac{\text{pJ}}{\text{m}}$$

Converting this continuum-model value into an atomic DMI strength [37] results in $d_\text{tot} = -(1.04 \pm 0.05)$ meV per bond, where the total DMI strength is referenced to the interface cobalt layer only (see Supplemental Material S1 [30]).

In conclusion, SEMPA is used to determine the strength of the Dzyaloshinskii-Moriya interaction of the epitaxially grown Co/Ir(111) interface. The growth and preparation conditions ensure an ideal interface between Co and Ir, making the experimentally determined DMI strength comparable to those from published *ab-initio* calculations. Below a thickness of 8.8 ML, purely Néel-oriented DW with a clockwise sense of rotation are observed. Above this thickness the stray-field contribution shifts the energy minimum towards a partial Bloch orientation of the DW, giving access to the strength of the DMI. Together with our earlier investigation of the DMI at the Co/Pt(111) interface [38], we verify the opposing signs of the DMI at the Co/Ir(111) and Co/Pt(111) interfaces at least for the ideal systems which is in agreement with the calculations and in contrast to the findings in [29, 33, 35, 36]. The determined DMI strength matches the value calculated in Perini *et al.* [8], but significantly deviates from the smaller and larger values given in [9, 19, 20].

**Acknowledgements**

We acknowledge financial support by the DFG within SFB 668. The authors would like to thank Sebastian Meyer, Bertrand Dupé, Levente Rózsa and Aurelien Manchon for fruitful discussions regarding the conversion between different DMI definitions, Christian Heyn for thickness calibration via atomic force microscopy, Kai Litzius for sharing a MicroMagnum version that includes the Dzyaloshinskii-Moriya interaction, Parisa Bayat for additional energy dispersive X-ray spectroscopy and Kirsten von Bergmann for information regarding the growth of cobalt on Ir(111).

# Supplemental Material

### S1. Conversion between different DMI calculations

*Ab initio* calculations offer the possibility to deduce the DMI strength from first principles. However, the comparison between the results from various groups and with micromagnetic continuum theory is not straight forward, as different units and definitions are in use. So, in the first place, a common description is required. Considering the DMI as an atomic interaction between nearest-neighbor spins $\vec{s}_i$ and $\vec{s}_j$, the DMI energy can be written as

$$E_{\text{DMI}} = \sum_{\langle i,j \rangle} \vec{d}_{ij} \cdot (\vec{s}_i \times \vec{s}_j) \tag{S1}$$

where the summation is over all bonds $\langle i,j \rangle$ between the nearest-neighbor atoms (each bond counted only once) and $\vec{d}_{ij}$ ist the corresponding atomic nearest-neighbor DMI vector. Breaking the symmetry in the $\vec{z}$-direction by an interface leads to a DMI vector $\vec{d}_{ij} = d(\vec{e}_z \times \vec{e}_{ij})$ with the unit vectors $\vec{e}_z$ and $\vec{e}_{ij}$ pointing along the $\vec{z}$-direction and from atom $i$ to atom $j$, respectively [1]. Thereby the DMI strength $d$ describes the interfacial DMI with a scalar quantity. In their calculation, Yang *et al.* [2] sum up the contributions from all monolayers, deducing a total DMI strength $d_{\text{tot}}$, that can be seen as the DMI strength concentrated in a single atomic layer producing an equivalent effect [2]. They calculate for 3 ML of cobalt on 3 ML of Ir(111) $d_{\text{tot}} = -0.40$ meV [3, 4].

As an interface contribution, the micromagnetic DMI strength in the thin-film limit can be treated using a volume-like term $D_s/t$ where $t$ is the thickness of the ferromagnetic material and $D_s$ the interfacial DMI constant, in analogy to the surface anisotropy $K_s$ [5, 6, 7, 8]. The micromagnetic energy per unit volume of magnetic material [9] reads in the continuum form

$$E_{\mu\text{-mag}} = D_s/t \left( m_z \operatorname{div} \vec{m} - (\vec{m} \cdot \vec{\nabla}) m_z \right), \tag{S2}$$

where $\vec{m}$ is the unit vector of the continuous magnetization. In the following, the individual results will be converted to this $D_s$ constant to enable a thickness-independent comparison. The relation between Eqs. (S1) and (S2) is derived in the supplemental material of [2], leading to

$$D_{\text{s}} = \frac{\sqrt{6}\, d_{\text{tot}}}{a_{\text{sub}}} \tag{S3}$$

with $a_{\text{sub}}$ the lattice constant of the substrate ($a_{\text{Ir}} = 3.839$ Å), assuming the overlayer grows pseudomorphically. From $d_{\text{tot}} = -0.40$ meV it follows $D_{\text{s}} = -0.42$ pJ/m for the value of Yang *et al.* [3, 4].

Belabbes *et al.* [10] calculate the energy of spin spirals with a wave vector $q$ for various monolayers of 3d materials. The energy, normalized per atom, is then fitted using the form $E_{\text{atom}} = A_{\text{atom}} q^2 + D_{\text{atom}} q + \frac{1}{2} K_{\text{atom}}$, where $A_{\text{atom}}$, $D_{\text{atom}}$ and $K_{\text{atom}}$ denote the spin stiffness, spiralization and effective anisotropy. Inserting $\vec{m} = \sin(q \cdot x)\,\vec{e}_x + \cos(q \cdot x)\,\vec{e}_z$ into Eq. (S2) leads for the micromagnetic energy (per volume unit) to

$$E_{\mu\text{-mag}} = D_{\text{s}}/t \; q \tag{S4}$$

For an fcc stacking, the DMI energy per atom ($E_{\text{atom}} = D_{\text{atom}} q$) is referring to an atomic volume of $\frac{1}{4} a_{\text{sub}}^3$, leading to:

$$E_{\mu\text{-mag}} = \frac{E_{\text{atom}}}{\frac{1}{4} a_{\text{sub}}^3} \tag{S5}$$

$$D_{\text{s}}/t = 4 \frac{D_{\text{atom}}}{a_{\text{sub}}^3} \tag{S6}$$

For one monolayer of cobalt on top of six layers of Ir(111) they obtain $D_{\text{atom}} = -3.05$ meV nm, leading to $D_{\text{s}} = -7.66$ pJ/m.

In a third calculation, Perini *et al.* [11] obtain a DMI strength per cobalt atom of $\tilde{d} = -0.54$ meV in 1 ML of cobalt on top of 9 ML of Ir(111). Although this value is close to the value of Yang *et al.* and appears in the same units, their definition is different, yielding a factor of two between both results. For their definition, Perini *et al.* use the energy per length of a domain wall in presence of the DMI $E_{\text{DW}} = \frac{8}{a_{\text{sub}}} \sqrt{|JK|} - \frac{2\pi\sqrt{6}}{a_{\text{sub}}} |\tilde{d}|$. As a 180° domain wall corresponds to a wave vector of $q = \pm\pi$ in Eq. (S4) one obtains:

$$-\pi \frac{D_{\text{s}}}{t} = -\frac{2\pi\sqrt{6}}{a_{\text{sub}}} \tilde{d} \frac{1}{t} \tag{S7}$$

$$D_{\text{s}} = \frac{2\sqrt{6}}{a_{\text{sub.}}} \tilde{d} \tag{S8}$$

We emphasize the factor 2 between Eqs. (S3) and (S8). Using $\tilde{d} = -0.54$ meV one obtains $D_{\text{s}} = -1.10$ pJ/m.

## S2. Details on the measurement and sample preparation

Magnetic domain walls in the Co/Ir(111) system were investigated by means of Scanning Electron Microscopy with Polarization Analysis (SEMPA). The utilized SEMPA instrument records both in-plane components of the magnetization simultaneously and allows for vectorial imaging of the orientation between both components with 4° precision for each pixel of an image [12]. Typical dwell times of 25 ms per pixel were used with a 6 keV/3.0 nA primary electron beam giving 3.5 million counts/s in single-electron detection. Furthermore, the line scan data of Fig. 2 are compiled from 24 individually realigned scans giving a total dwell time of 5.9 s/pixel. The horizontal and vertical components were recorded simultaneously which excludes any shift between both independent directions.

Domains and domain walls were imaged over the whole stated thickness range. The main challenge in the experiment was to record images with a quality good enough, not only to see domain walls but also to analyze their orientation. From a practical point of view, focusing an electron beam on a nearly ideal single-crystal surface is challenging, as defects or dirt on the surface are needed to adjust the optics. Therefore, the maximum performance of the instrument was not reached and for example, in the case of Fig. 2, domain walls with a width of 14.7 nm were imaged using a resolution of 70 nm. Towards lower thicknesses, the domain-wall width decreases further and thus makes the recording even more challenging. Hence, domain walls were imaged at lower thicknesses, but the obtained micrographs do not allow for a proper analysis of the angles. In contrast, more data points were obtained towards the SRT, as the domain walls, which become wider in the vicinity of the SRT, are easier to image.

The Co/Ir(111) sample was prepared in a preparation chamber (base pressure of $5 \times 10^{-11}$ mbar) that is directly attached to the SEMPA instrument. The as-polished Ir(111) single crystal (miscut below 0.1°, corresponding to a minimum terrace width of 130 nm) was periodically flashed (duration of 1 min., period of 20 min.) at 1400 K in oxygen atmosphere ($10^{-8}$ mbar) for 3 days. Subsequent annealing cycles at 1550 K without oxygen were carried out to remove the oxide. Afterwards a clean, atomically flat Ir(111) surface was prepared by argon ion sputtering (1 µA/500 eV) and simultaneous annealing at 1400 K in cycles of 2 min duration and 20 min. period.

In FIG. S1 a differentiated Auger electron spectrum from the clean Ir(111) surface is shown. Blue and gray curves show the measured spectrum and a reference spectrum of iridium [13] at 3 kV beam energy, respectively. The energies of the most frequent C- and O-contaminants are marked in red and show no peaks, which demonstrates the cleanliness of the surface.

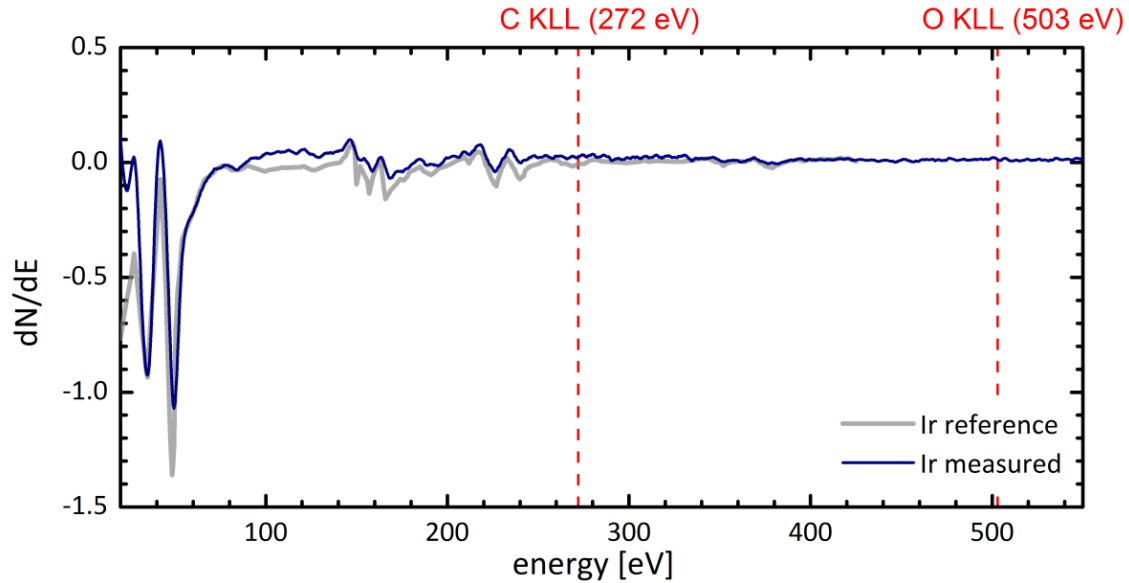

*FIG. S1. Differentiated Auger electron spectrum of the clean Ir(111) single crystal. Blue and gray curve show the measured spectrum and a reference spectrum from [13], respectively. The energies of the most frequent C- and O-adsorbates are marked in red. No contaminants can be identified in the recorded data.*

In FIG. S2(a) a sharp p(1 × 1) pattern of low-energy electron diffraction (LEED) from the clean Ir(111) single crystal is shown, indicating a long-range crystallographic order of the surface without superstructures. FIG. S2(b) displays the LEED pattern at $(1.45 \pm 0.03)$ ML average cobalt coverage on the Ir(111) surface. The threefold symmetry indicates a fixed stacking of the cobalt atoms of either fcc or hcp with respect to the substrate. Furthermore, the same lattice constants visible in both LEED patterns prove a pseudomorphically grown first cobalt layer. On further increase of cobalt thickness the lattice starts to reduce strain and shrinks towards the bulk cobalt lattice constant. FIG. S3 shows the change of in-plane lattice constant after deposition of (17.7 ± 0.4) ML cobalt. Using the mentioned shutter, half of the Ir(111) surface is covered with (17.7 ± 0.4) ML of cobalt, while the other half remains clean. The sample is then positioned in a way that the LEED beam illuminates similar sizes of covered and uncovered areas, giving LEED spots of both lattices simultaneously. The radius of the inner spots (corresponding to Ir(111)) is 7.4% smaller than the outer one, indicating a fully relaxed cobalt layer.

The thickness of ferromagnetic material is the most crucial parameter when measuring the interfacial DMI constant $D_\text{s}$, as the obtained value scales quadratically with thickness. For this reason, a precise knowledge of the evaporated thickness is essential. The cobalt wedge was evaporated at RT by electron-beam heating from a high-purity rod using the straight blade of a piezo-motor-controlled shutter at short distance from the surface (~1 mm). The deposition rate was kept constant by a feedback control of the cobalt ion flux and additionally, the sample current was monitored. During deposition the pressure did not rise above $2 \cdot 10^{-10}$ mbar. The wedge-shaped sample was then investigated by SEMPA in the as-grown state.

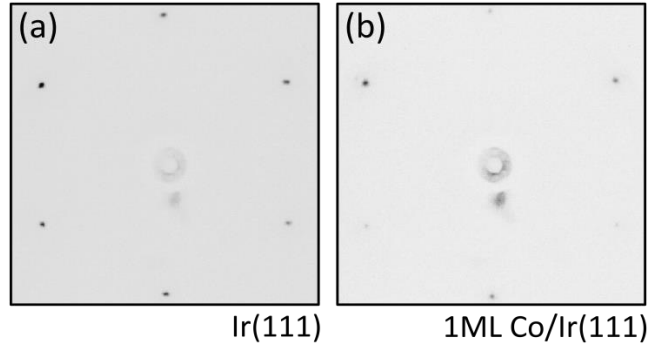

Ir(111)    1ML Co/Ir(111)

FIG. S2. *LEED patterns from the clean Ir(111) single crystal (a) and after deposition of 1.45 ML cobalt (b). Both patterns were recorded at a beam energy of 50 eV and show sharp LEED-spots at the same positions without superstructure.*

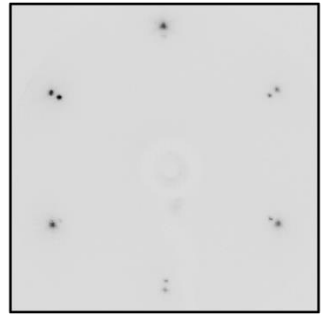

FIG. S3. *Low-energy electron diffraction (LEED) of the Ir(111) single crystal at 70 eV. Only half of the Ir(111) surface is covered with $(\mathbf{17.7 \pm 0.4})$ M of cobalt using a half-closed shutter. The LEED beam illuminates covered and uncovered parts showing the LEED spots of both lattices. The inner lattice (corresponding to Ir(111)) is 7.4% smaller than the outer one indicating a fully relaxed cobalt layer.*

To calibrate the deposition rate, a sharp cobalt step was evaporated onto a Pt/Si$_3$N$_4$/Si sample by means of a shadow mask and measured *ex situ* by atomic force microscopy (AFM). Evaporating cobalt directly by electron-beam heating onto Si$_3$N$_4$ resulted in a very inhomogeneous cobalt deposition, probably due to sample charging. Prior *ex situ* coverage of the Si$_3$N$_4$ film with 10 nm of platinum establishes conductivity and yields homogeneous growth of the cobalt. To prevent oxidation of the cobalt layer, which is connected with a thickness increase [14], the whole sample was covered *in situ* by a (1.5 ± 0.1) nm platinum layer using electron-beam deposition. A fit to the obtained AFM height profiles yields a step height of (9.6 ± 0.2) nm and a width of $2\sigma =$ (315 ± 3) nm. The given accuracy is derived by calibration against a standard (HS-20MG from Ted Pella, Inc.) providing 20.0 nm steps with 2% stated accuracy. An evaporation rate of (1.066 ± 0.022) Å/min. follows. Using the monolayer distance of bulk cobalt in the fcc state (2.0464 Å in the <111>-direction) this corresponds to (0.5209 ± 0.011) ML/min.

We want to point out that the calibration via an evaporation rate gives the amount of evaporated cobalt atoms per unit area, which is not necessarily identical to the physical film thickness, when the strained growth of the first ML is considered. In the performed analysis, the interfacial DMI energy of the DW is counterbalanced by the dipolar energy of the DW charge, which depends directly on the magnetic moment per unit area and not on the actual thickness of the cobalt layer. For this reason, our DMI analysis is independent of potential volume changes due to the strained growth of the initial layers, as long as the magnetic moments of Co are preserved. The given thicknesses can thus be considered as effective "magnetic" thicknesses.

### S3. Analysis of the domain-wall profiles measured by SEMPA

In the following, a domain-wall profile of the common Bloch form [15] is assumed:

$$\mathrm{Asy}(x) = \frac{\mathrm{Asy}_{\mathrm{max}}}{\cosh\left(\frac{x}{\Delta}\right)} \tag{S9}$$

$\mathrm{Asy}_{\mathrm{max}}$ denotes the intensity asymmetry from the polarization analyzer, that corresponds to the spin polarization of the secondary electrons originating from a surface area where the magnetization is fully aligned with the sensitive axis of the analyzer. $\Delta$ is the domain-wall width parameter that is given by $\Delta = \sqrt{\frac{A}{K_{\mathrm{eff}}}}$, with A the exchange stiffness and $K_{\mathrm{eff}}$ the effective anisotropy constant. Originally derived for Bloch walls, this domain-wall profile is still a good assumption for the Néel-oriented walls, that are induced by DMI [16]. On the one hand the DMI does not directly affect the wall width but primarily the in-plane rotational angle. On the other hand the tiny correction of the domain-wall width predicted by simulation is only a secondary effect of the magnetic charges in the Néel-oriented wall compared to the uncharged Bloch wall [9, 17] and can be neglected.

However, the widths of the measured domain wall profiles do not reproduce the actual domain-wall width parameter $\Delta$, but they are a consequence of imperfect focusing of the scanning electron beam, as it is obvious from the strongly reduced peak asymmetry value [18]. The actual domain-wall profile is convoluted by the profile of the scanning electron beam, which leads to a considerable widening of the apparent domain-wall profile that is accompanied by a loss in contrast. From the knowledge of the maximum contrast $\mathrm{Asy}_{\mathrm{max}}$ the actual domain-wall width can be reconstructed following the description in [18]. The measured domain-wall profile is fitted with a Gaussian profile

$$\overrightarrow{\mathrm{Asy}}(x) = \begin{pmatrix} V_{\mathrm{horz.}} \\ V_{\mathrm{vert.}} \end{pmatrix} \frac{1}{\sqrt{2\pi}\cdot\sigma} \exp\left[-\frac{1}{2}\left(\frac{x-x_0}{\sigma}\right)^2\right]. \tag{10}$$

While the integrated intensities of the horizontal ($V_{\text{horz.}}$) and vertical component ($V_{\text{vert.}}$) are independent parameters, the standard deviation (or the width) $\sigma$ and the position value $x_0$ are the same for both components.

The integral over the measured profile $V = \sqrt{V_{\text{horz.}}^2 + V_{\text{vert.}}^2}$ is independent of the actual beam width and is equal to the integral over the ideal domain-wall profile $\text{Asy}_{\text{max}} \pi \Delta$. Following the domain-wall-width definition by Lilley ($d_{\text{w}} = \pi \Delta$) [19] one finds

$$d_{\text{w}} = \frac{V}{\text{Asy}_{\text{max}}}. \tag{11}$$

The maximum asymmetry of the spin polarization $\text{Asy}_{\text{max}} = (4.43 \pm 0.03)\%$ is measured on an in-plane 17 ML thick cobalt film. Accounting for an information depth of 0.5 nm [20] this value has been scaled to 9.0 ML film thickness (correction factor 0.976). Therefore, one obtains for the shown domain-wall profiles at 9.0 ML:

$$d_{\text{w}} = \frac{(0.63 \pm 0.03)\text{nm}}{(4.43 \pm 0.03)\% \cdot 0.976} = (14.6 \pm 0.7)\text{nm} \tag{12}$$

### S4. Estimating the DMI strength with an analytical approximation

The interfacial DMI favors Néel-oriented DWs with a fixed sense of rotation. Its contribution to the domain-wall energy $\sigma$ is given by

$$\sigma_{\text{D}} = \pi \frac{D_{\text{s}}}{t} \cos \Phi. \tag{S133}$$

Minimization of $\sigma_{\text{D}}$ implies that a DW of clockwise (counterclockwise) sense of rotation is associated with a negative (positive) sign of the DMI. In contrast, the minimization of the stray-field energy favors Bloch walls avoiding magnetic volume charges. As the stray-field energy of a DW scales quadratically with the induced volume charges, it can be expressed by [21, 9, 22]

$$\sigma_{\text{stray}} = \sigma_{\text{NB}} \cos^2 \Phi. \tag{S144}$$

Here $\sigma_{\mathrm{NB}}$ denotes the difference in stay-field energy between a pure Néel and a pure Bloch DW. It depends on the thickness $t$ and the saturation magnetization $M_s$ of the cobalt layer ($M_s = 1440$ kA/m [23]) and can be obtained numerically as in [24, 25, 26, 27] or in an analytical approximation as calculated by Tarasenko *et al.* [21] or more accurately by Lemesh *et al.* [22]. The latter interpolate their solutions for ultra-thin films (as beforehand determined by Tarasenko *et al.* [21]) and thick films. Minimizing the energy of a DW, they obtain in a second step:

$$\cos \Phi = \begin{cases} -D_s/D_{\mathrm{thr}} & \forall\, |D_s| < D_{\mathrm{thr}} \\ -\mathrm{sgn}(D_s) & \forall\, |D_s| \geq D_{\mathrm{thr}} \end{cases} \qquad D_{\mathrm{thr}} := \frac{2\mu_0 M_s^2\, t^2}{\frac{\pi^2}{\ln(2)} + \pi\, t\sqrt{\frac{K_u + \frac{\mu_0 M_s^2}{2}}{A}}} \quad (S155)$$

Thus, above a certain threshold DMI $D_{\mathrm{thr}}$ of the magnetic film, purely Néel-oriented DWs occur, giving access to the sign and a lower limit of the DMI. For $|D_s| < D_{\mathrm{thr}}$ a smooth transition from Néel- to Bloch-oriented DWs takes place and the angle of the DW magnetization $\Phi$ can be used as a direct probe for $D_s$. The threshold DMI $D_{\mathrm{thr}}$ dependents on the thickness of the cobalt layer $t$, the saturation magnetization $M_s$ of Co, and implicitly on the exchange stiffness $A$ and the uniaxial anisotropy constant $K_u$. Although the energy minimization involves only the DMI and the stray-field energy, the threshold DMI $D_{\mathrm{thr}}$ shows a dependence on $A$ and $K_u$. This dependence arises, as in the thick-film limit the DW width of a purely Néel-oriented DW, which in this limit is given by $\pi\sqrt{\left(K_u + \frac{\mu_0 M_s^2}{2}\right)/A}$, contributes to the stray-field energy. Apparently, the left and right term under the fraction line in Eq. 15 represent the mentioned ultra-thin and thick film limits, respectively. However, in the present study film thicknesses below 9 ML (1.84 nm) are investigated and thus the dependence on the actual domain-wall width can be considered as a tiny correction (i.e., doubling the DW width by reducing the anisotropy accordingly results only in an overestimation of the DMI by 1.8%, which is still below our experimental accuracy of 4%). This is consistent with the findings of numerical calculations [24, 25, 9] that report practically no dependence of the DMI on the actual domain-wall width. Indeed, the DMI term (Eq. 13) does not directly affect the wall width but primarily the in-plane rotational angle. Magnetic volume charges resulting from a Néel-oriented wall instead of an uncharged Bloch wall result only in a small correction of the DW width as a secondary effect [9, 17, 22].

Using the previously performed analysis of the domain-wall profiles (Supplemental Material S3), a uniaxial anisotropy constant of $K_u = (2.5 \pm 0.2) \cdot 10^6$ J/m$^3$ can be deduced from the measured DW width of $(14.6 \pm 0.7)$ nm at 9.0 ML. As the domain-wall width $d_w$ is defined by the exchange stiffness $A$ and an effective anisotropy $K_{eff}$ (i.e. $d_w = \pi\sqrt{A/K_{eff}}$) [15, 18, 22], the known domain-wall width and exchange stiffness conclude an effective anisotropy of $K_{eff} = (1.3 \pm 0.2) \cdot 10^6$ J/m$^3$. Here the approach in [18, 28] was followed to scale the bulk value of the exchange stiffness linearly with the average coordination number of the Co atoms giving $A = 28.3 \cdot 10^{-12}$ J/m at 9.0 ML. As this thickness dependence of $A$ is quite low at the investigated coverages of cobalt, a constant value of $28.3 \cdot 10^{-12}$ J/m is used. In a reasonable approximation the uniaxial anisotropy constant $K_u$ can be obtained from the effective anisotropy $K_{eff}$ via $K_u \approx K_{eff} + \frac{\mu_0}{2}M_s^2 = (2.6 \pm 0.2)10^6$ J/m$^3$. However, considering the shape anisotropy by $\frac{\mu_0}{2}M_s^2$ results in a slight overestimation, especially for thin DWs or thick films. Using micromagnetic simulation (see Supplemental Material S5) we find that the measured domain-wall width at 9 ML coverage of cobalt corresponds to $K_u = (2.5 \pm 0.2) \cdot 10^6$ J/m$^3$.

As the uniaxial anisotropy as well as the exchange stiffness are of minor relevance in Eq. (S155), it is sufficient to (S133)approximate the thickness dependence of the uniaxial anisotropy below 9 ML by using $K_u = K_s/t$, with a surface anisotropy of $K_s = (4.5 \pm 0.3) \cdot 10^{-3}$ J/m$^2$. This neglects a possible additional volume anisotropy $K_v$ in the thickness dependence of the uniaxial anisotropy $K_u$, implying that the cobalt grows fcc and without lattice distortion over this thickness regime (i.e. between 8.5 ML and 9 ML), which is probably not the case. The second cobalt monolayer already starts to relieve the strain between the cobalt and the iridium lattice and therefore exhibits a reconstruction pattern containing both hcp as well as fcc areas [29]. The third and the fourth layer of cobalt further proceed in reducing the strain towards an undistorted lattice [29] and hence, we have to consider a mixed stacking at the relevant thickness around 9 ML. However, an hcp stacking goes along with a volume anisotropy of $K_v = 5 \cdot 10^5 \text{ J/m}^3$ [23]. Therefore the remaining uncertainly in the uniaxial anisotropy is in a worst-case estimation $\Delta K_u = 5 \cdot 10^5 \text{ J/m}^3 \cdot (1.84 \text{ nm}/t - 1)$, which we take into account in the final error estimate. Nevertheless, due to the mentioned weak dependence on the uniaxial anisotropy this source of error is insignificant. Furthermore, the effect of strain relief at lower coverages of cobalt enters in the determined surface anisotropy $K_s$, which explains why a value higher that of a monolayer of cobalt is observed [30].

The equation of Lemesh *et al.* [22] has been fitted to the experimental data with $D_s$ as the only fit parameter. This fit yields $D_s = -(1.04 \pm 0.01)\frac{\text{pJ}}{\text{m}}$. We are not able to estimate the reliability of the assumptions made in deriving the analytical approximation, where solutions for ultra-thin films (accounting only for surface charges) and ultra-thick films (accounting only for volume charges) are interpolated. This makes a reasonable error handling challenging and hence this ansatz is only considered as an additional, successful crosscheck for the following DMI quantification using micromagnetic simulations.

## S5. Quantifying the DMI strength with micromagnetic simulations

Micromagnetic simulations are performed using a modified version of MicroMagnum [31] that accounts for the Dzyaloshinskii-Moriya interaction [32]. Stripe domains have been relaxed (until the maximum change of the magnetization angle is below $0.01\,\text{deg./ns}$) in a periodic mesh (infinite periodicity in both in-plane directions) of various thicknesses (no periodicity out of plane). The infinite periodicity in both in-plane direction avoids unwanted effects from magnetic edge charges as discussed in [33, 34]. The simulated mesh has a size of 500 nm perpendicular (x-direction) and 50 nm parallel (y-direction) to the stripe domains. Cell sizes of 0.5 nm and 1 nm are used in x-direction and y-direction, respectively, ensuring that the DWs are simulated in a gradual transition with 0.5 nm steps.

Likewise to Supplemental Material S4, a saturation magnetization of 1440 kA/m and an exchange stiffness of $A = 28.3 \cdot 10^{-12}\,\text{J/m}$ have been used. For the micromagnetic DMI and uniaxial anisotropy $K_\text{u}$ an $1/t$-dependency applies, hence they have been scaled like $D_\text{s}/t$ and $K_\text{u} = K_\text{s}/t$ with $K_\text{s} = 4.5 \cdot 10^{-3}\,\text{J/m}^2$ on varying the thickness of the magnetic film $t$. This surface anisotropy results in combination with the exchange constant of $A = 28.3 \cdot 10^{-12}\,\text{J/m}$ in the experimentally observed DW width of $(14.6 \pm 0.7)$ nm (see eq. S12) ensuring a reasonable calculation of the stray-field energy in the DW. A tiny uncertainty due to a small volume anisotropy has been made subject of discussion in Supplemental Material S4 and is included in the error estimate. Furthermore, we emphasize that the actual exchange stiffness as well as the anisotropy are irrelevant in the performed analysis as long as the domain-wall widths are correct, which has been measured at 9 ML.

The only free parameter $D_\text{s}$ is obtained by matching simulated and experimental angles of DW magnetization iteratively. We find that fitting the simulated angles with (S155) results in a 2.8% to low DMI strength. Scaling the strength appropriately in a way that the fit on the simulated curve and experimental data coincide gives a DMI strength of $D_s = -1.065\,\text{pJ/m}$. Together with further simulations using $D_s = -1.07\,\frac{\text{pJ}}{\text{m}}$, the statistical error of the experimental data points is assessed, giving $D_s = -(1.07 \pm 0.01)\,\frac{\text{pJ}}{\text{m}}$.

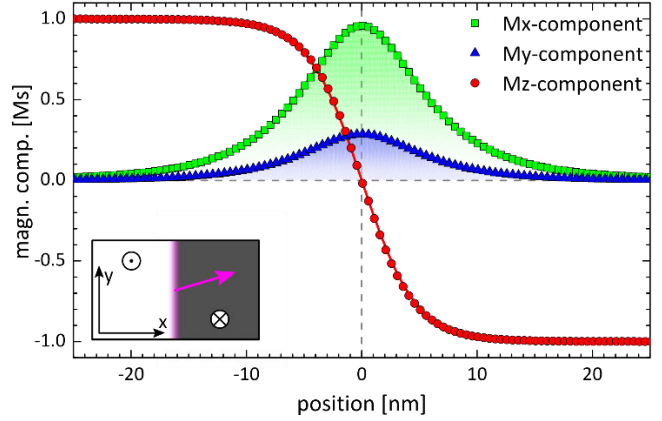

FIG. S4. Simulated domain-wall profile of 9 ML of cobalt. An interfacial DMI constant of $D_s = -1.065\,\text{pJ/m}$ results in an angle of DW magnetization of $16.4°$ that is consistent with the experiment.

Fig. S4. shows the profile of a simulated DW at a cobalt thickness of 9 ML (corresponding to 1.84 nm) with the interfacial DMI constant of $D_s = -1.065\,\text{pJ/m}$. The profile along the x-direction cuts the DW that runs along the y-direction perpendicularly (see inset in the lower left corner). Therefore, the red dots show the z-component of magnetization, which is oriented out-of-plane. The green squares visualize the x-component of magnetization corresponding to the Néel content of the DW, and the blue triangles the y-component corresponding to the Bloch content of the DW. The angle of the DW magnetization $\phi$ as well as the DW width $d_w = \pi \cdot \Delta$ is determined in a single fitting procedure using a profile of the common Bloch form (i.e. $\tanh(x/\Delta)$-profile for the out-of-plane component of the magnetization and $1/\cosh(x/\Delta)$-profile for the y-component of the magnetization [15]). Together with an additional rotation matrix that rotates the in-plane magnetization by the angle of DW magnetization $\phi$ one obtains:

$$\vec{M} = M_s \cdot \begin{pmatrix} \sin(\phi) & \cos(\phi) & 0 \\ -\cos(\phi) & \sin(\phi) & 0 \\ 0 & 0 & 1 \end{pmatrix} \cdot \begin{pmatrix} 0 \\ 1/\cosh(x/\Delta) \\ -\tanh(x/\Delta) \end{pmatrix} \qquad (16)$$

Fitting Eq. 16 to the simulated DW profile in Fig. S4. results in $\phi = (16.38 \pm 0.05)°$ and $d_\text{w} = (14.7 \pm 0.1)$ nm, both of which are in agreement with the experimentally obtained data on the same thickness, as shown in Fig. 2.